\documentclass[12pt]{article}
\usepackage{graphics,graphicx}
\usepackage{amssymb,epsfig,amsmath,euscript,array}
\usepackage{cite}
\usepackage{pstricks}
\usepackage{color}

\makeatletter
\@addtoreset{equation}{section}
\makeatother

\newcounter{multieqs}

\newcommand{\be}{\begin{equation}}
\newcommand{\ee}{\end{equation}}

\newcommand{\bm}[1]{\mbox{\boldmath $#1$}}

\newcommand{\kslash}{k \!\!\! / }

\newcommand{\lslash}{l \!\! / }
\newcommand{\Pslash}{P \!\!\!\! / }

\newcommand{\islash}{i \!\!\! / }
\newcommand{\jslash}{j \!\!\! / }
\newcommand{\aslash}{a \!\!\! / }
\newcommand{\bslash}{{b \hspace{-6pt} \slash} }

\newcommand{\onslash}{1 \!\!\! / }
\newcommand{\twslash}{2 \!\!\!/ }
\newcommand{\thslash}{3 \!\!\!/ }
\newcommand{\foslash}{4 \!\!\! / }
\newcommand{\fislash}{5 \!\!\! / }

\newcommand{\mslash}{m \!\!\! / }

\def\bd{\begin{document}}
\def\ed{\end{document}}
\def\nn{\nonumber}
\def\bea{\begin{eqnarray}}
\def\eea{\end{eqnarray}}

\def\ab{(ijab)}
\def\ba{(ijba)}
\def\ijab{{\tr}_{+}(\islash\, \jslash\, \aslash \, \bslash)}
\def\ijba{{\tr}_{+}(\islash\, \jslash\, \bslash \, \aslash)}
\def\ijaP{{\tr}_{+}(\islash\, \jslash\, \aslash \, \Pslash)}
\def\ijPLa{{\tr}_{+}(\islash\, \jslash\, \Pslash_L \, \aslash)}
\def\ijaPL{{\tr}_{+}(\islash\, \jslash\, \aslash \, \Pslash_L)}
\def\ijPLza{{\tr}_{+}(\islash\, \jslash\, \Pslash_{L;z} \, \aslash)}
\def\ijaPLz{{\tr}_{+}(\islash\, \jslash\, \aslash \, \Pslash_{L;z})}
\def\ijPa{{\tr}_{+}(\islash\, \jslash\, \Pslash \, \aslash)}
\def\iaPb{{\tr}_{+}(\islash\, \aslash\, \Pslash \, \bslash)}
\def\ibPa{{\tr}_{+}(\islash\, \bslash\, \Pslash \, \aslash)}
\def\ijPmu{{\tr}_{+}(\islash\, \jslash\, \Pslash \, \mu)}
\def\ibmuP{{\tr}_{+}(\islash\, \bslash\, \mu \, \Pslash)}
\def\ibmua{{\tr}_{+}(\islash\, \bslash\, \mu \, \aslash)}
\def\iamub{{\tr}_{+}(\islash\, \aslash\, \mu \, \bslash)}
\def\jaPb{{\tr}_{+}(\jslash\, \aslash\, \Pslash \, \bslash)}
\def\ijmuP{{\tr}_{+}(\islash\, \jslash\, \mu \, \Pslash)}
\def\ijmum{{\tr}_{+}(\islash\, \jslash\, \mu \, \mslash)}
\def\ijmmu{{\tr}_{+}(\islash\, \jslash\, \mslash \, \mu)}
\def\ijmP{{\tr}_{+}(\islash\, \jslash\, \mslash \, \Pslash)}
\def\iabP{{\tr}_{+}(\islash\, \aslash\, \bslash \, \Pslash)}
\def\ijbP{{\tr}_{+}(\islash\, \jslash\, \bslash \, \Pslash)}
\def\jbPa{{\tr}_{+}(\jslash\, \bslash\, \Pslash \, \aslash)}
\def\ijPb{{\tr}_{+}(\islash\, \jslash\, \Pslash \, \bslash)}
\def\jbmua{{\tr}_{+}(\jslash\, \bslash\, \mu \, \aslash)}

\def\loablt{ {\tr}_{+}(\lslash_1\, \aslash \, \bslash\, \lslash_2)}

\def\ijlolt{{\tr}_{+}(\islash\, \jslash\, \lslash_1 \, \lslash_2)}
\def\ijltlo{{\tr}_{+}(\islash\, \jslash\, \lslash_2 \, \lslash_1)}
\def\ibloa{{\tr}_{+}(\islash\, \bslash\, \lslash_1 \, \aslash)}
\def\jaltb{{\tr}_{+}(\jslash\, \aslash\, \lslash_2 \, \bslash)}
\def\ialtb{{\tr}_{+}(\islash\, \aslash\, \lslash_2 \, \bslash)}
\def\bltloa{{\tr}_{+}(\bslash\, \lslash_2\, \lslash_1 \, \aslash)}
\def\jbloa{{\tr}_{+}(\jslash\, \bslash\, \lslash_1 \, \aslash)}
\def\ibPb{{\tr}_{+}(\islash\, \bslash\, \Pslash \, \bslash)}
\def\ijltb{{\tr}_{+}(\islash\, \jslash\, \lslash_2 \, \bslash)}

\def\ijloa{{\tr}_{+}(\islash\, \jslash\,  \lslash_1 \, \aslash)}
\def\ijblt{{\tr}_{+}(\islash\, \jslash\,  \bslash \, \lslash_2)}

\def\jakb{{\tr}_{+}(\jslash\, \aslash\, \kslash \, \bslash)}
\def\iakb{{\tr}_{+}(\islash\, \aslash\, \kslash \, \bslash)}

\def\tofo{{\tr}_{+}(\onslash\, \thslash\, \twslash \, \foslash)}
\def\foto{{\tr}_{+}(\onslash\, \thslash\, \foslash \, \twslash)}
\def\tofi{{\tr}_{+}(\onslash\, \thslash\, \twslash \, \fislash)}
\def\fito{{\tr}_{+}(\onslash\, \thslash\, \fislash \, \twslash)}

\def\lrangle#1#2{\langle #1\,#2\rangle}

\def\Li{{$\rm Li}_2$}
\def\eps{\epsilon}
\def\epsuv{{\epsilon_{\rm \mbox{\tiny UV}}}}
\let\bm=\bibitem
\let\la=\label

\def\npb#1#2#3{Nucl. Phys. {\bf{B#1}} #3 (#2)}
\def\plb#1#2#3{Phys. Lett. {\bf{#1B}} #3 (#2)}
\def\prl#1#2#3{Phys. Rev. Lett. {\bf{#1}} #3 (#2)}
\def\prd#1#2#3{Phys. Rev. {D \bf{#1}} #3 (#2)}
\def\cmp#1#2#3{Comm. Math. Phys. {\bf{#1}} #3 (#2)}
\def\cqg#1#2#3{Class. Quantum Grav. {\bf{#1}} #3 (#2)}
\def\nppsa#1#2#3{Nucl. Phys. B (Proc. Suppl.) {\bf{#1A}}#3 (#2)}
\def\ap#1#2#3{Ann. of Phys. {\bf{#1}} #3 (#2)}
\def\ijmp#1#2#3{Int. J. Mod. Phys. {\bf{A#1}} #3 (#2)}
\def\rmp#1#2#3{Rev. Mod. Phys. {\bf{#1}} #3 (#2)}
\def\mpla#1#2#3{Mod. Phys. Lett. {\bf A#1} #3 (#2)}
\def\jhep#1#2#3{J. High Energy Phys. {\bf #1} #3 (#2)}
\def\atmp#1#2#3{Adv. Theor. Math. Phys. {\bf #1} #3 (#2)}
\newcommand{\EQ}[1]{\begin{equation} #1 \end{equation}}
\newcommand{\AL}[1]{\begin{subequations}\begin{align} #1 \end{align}\end{subequations}}
\newcommand{\SP}[1]{\begin{equation}\begin{split} #1 \end{split}\end{equation}}
\newcommand{\ALAT}[2]{\begin{subequations}\begin{alignat}{#1} #2 \end{alignat}
                        \end{subequations}}
\def\beqa{\begin{eqnarray}}
\def\eeqa{\end{eqnarray}}
\def\beq{\begin{equation}}
\def\eeq{\end{equation}}
\def\sst{\scriptscriptstyle}
\def\thetabar{\bar\theta}
\def\Tr{{\rm Tr}}
\def\one{\mbox{1 \kern-.59em {\rm l}}}
 \def\Nh{\hat{N}}

\newcommand{\half}{{\textstyle {1 \over 2}}}

\def\a{\alpha}      \def\da{{\dot\alpha}}
\def\b{\beta}       \def\db{{\dot\beta}}
\def\c{\gamma}  \def\G{\Gamma}  \def\cdt{\dot\gamma}
\def\d{\delta}  \def\D{\Delta}  \def\ddt{\dot\delta}
\def\e{\epsilon}        \def\vare{\varepsilon}
\def\f{\phi}    \def\F{\Phi}    \def\vvf{\f}
\def\h{\eta}
\def\k{\kappa}
\def\l{\lambda} \def\L{\Lambda}
\def\m{\mu} \def\n{\nu}
\def\o{\omega}
\def\p{\pi} \def\P{\Pi}
\def\r{\rho}
\def\s{\sigma}  \def\S{\Sigma}
\def\t{\tau}
\def\th{\theta} \def\Th{\Theta} \def\vth{\vartheta}
\def\X{\Xeta}
\def\z{\zeta}
\def\de{\partial}

\def\cA{{\cal A}} \def\cB{{\cal B}} \def\cC{{\cal C}}
\def\cD{{\cal D}} \def\cE{{\cal E}} \def\cF{{\cal F}}
\def\cG{{\cal G}} \def\cH{{\cal H}} \def\cI{{\cal I}}
\def\cJ{{\cal J}} \def\cK{{\cal K}} \def\cL{{\cal L}}
\def\cM{{\cal M}} \def\cN{{\cal N}} \def\cO{{\cal O}}
\def\cP{{\cal P}} \def\cQ{{\cal Q}} \def\cR{{\cal R}}
\def\cS{{\cal S}} \def\cT{{\cal T}} \def\cU{{\cal U}}
\def\cV{{\cal V}} \def\cW{{\cal W}} \def\cX{{\cal X}}
\def\cY{{\cal Y}} \def\cZ{{\cal Z}}

\def\ua{\underline{\alpha}}
\def\ub{\underline{\phantom{\alpha}}\!\!\!\beta}
\def\uc{\underline{\phantom{\alpha}}\!\!\!\gamma}
\def\um{\underline{\mu}}
\def\ud{\underline\delta}
\def\ue{\underline\epsilon}
\def\una{\underline a}\def\unA{\underline A}
\def\unb{\underline b}\def\unB{\underline B}
\def\unc{\underline c}\def\unC{\underline C}
\def\und{\underline d}\def\unD{\underline D}
\def\une{\underline e}\def\unE{\underline E}
\def\unf{\underline{\phantom{e}}\!\!\!\! f}\def\unF{\underline F}
\def\unm{\underline m}\def\unM{\underline M}
\def\unn{\underline n}\def\unN{\underline N}
\def\unp{\underline{\phantom{a}}\!\!\! p}\def\unP{\underline P}
\def\unq{\underline{\phantom{a}}\!\!\! q}
\def\unQ{\underline{\phantom{A}}\!\!\!\! Q}
\def\unH{\underline{H}}

\def\As {{A \hspace{-6.4pt} \slash}\;}
\def\bs {{b \hspace{-6.4pt} \slash}\;}
\def\Ds {{D \hspace{-6.4pt} \slash}\;}
\def\ds {{\del \hspace{-6.4pt} \slash}\;}
\def\ss {{\s \hspace{-6.4pt} \slash}\;}
\def\ks {{ k \hspace{-6.4pt} \slash}\;}
\def\ps {{p \hspace{-6.4pt} \slash}\;}
\def\pas {{{p_1} \hspace{-6.4pt} \slash}\;}
\def\pbs {{{p_2} \hspace{-6.4pt} \slash}\;}
\def\Ps {{P \hspace{-6.4pt} \slash}\;}
\def\Qs {{Q \hspace{-6.4pt} \slash}\;}

\def\Fh{\hat{F}}
\def\Vh{\hat{V}}
\def\Xh{\hat{X}}
\def\ah{\hat{a}}
\def\xh{\hat{x}}
\def\yh{\hat{y}}
\def\ph{\hat{p}}
\def\xih{\hat{\xi}}
\def\psit{\tilde{\psi}}
\def\Psit{\tilde{\Psi}}
\def\tht{\tilde{\th}}
\def\lt{\tilde{\lambda}}
\def\hl{\hat{\lambda}}
\def\hlt{\hat{\tilde{\lambda}}}
\def\llt{\tilde{l}}
\def\At{\tilde{A}}
\def\Qt{\tilde{Q}}
\def\Rt{\tilde{R}}
\def\Nt{\tilde{N}}

\def\at{\tilde{a}}
\def\st{\tilde{s}}
\def\ft{\tilde{f}}
\def\pt{\tilde{p}}
\def\qt{\tilde{q}}
\def\vt{\tilde{v}}
\def\nt{\tilde{n}}

\def\delb{\bar{\partial}}
\def\bz{\bar{z}}
\def\bD{\bar{D}}
\def\bB{\bar{B}}

\def\bk{{\bf k}}
\def\bl{{\bf l}}
\def\bp{{\bf p}}
\def\bq{{\bf q}}
\def\br{{\bf r}}
\def\bx{{\bf x}}
\def\by{{\bf y}}
\def\bR{{\bf R}}
\def\bV{{\bf V}}

\def\d{\delta}\def\D{\Delta}\def\ddt{\dot\delta}
\def\pa{\partial} \def\del{\partial}
\def\xx{\times}
\def\uno{\mbox{1 \kern-.59em {\rm l}}}
\def\trp{^{\top}}
\def\inv{^{-1}}
\def\dag{{^{\dagger}}}
\def\pr{^{\prime}}
\def\lan{\langle}
\def\ran{\rangle}
\def\rar{\rightarrow}
\def\lar{\leftarrow}
\def\lrar{\leftrightarrow}
\newcommand{\0}{\,\!}      
\def\one{1\!\!1\,\,}
\def\im{\imath}
\def\jm{\jmath}
\newcommand{\tr}{\mbox{tr}}
\newcommand{\slsh}[1]{/ \!\!\!\! #1}
\def\vac{|0\rangle}
\def\lvac{\langle 0|}
\def\hlf{\frac{1}{2}}
\def\ove#1{\frac{1}{#1}}
\def\Box{\square}
\def\ZZ{\mathbb{Z}}
\def\CC#1{({\bf #1})}
\def\bcomment#1{}
\def\bfhat#1{{\bf \hat{#1}}}
\def\VEV#1{\left\langle #1\right\rangle}
\newcommand{\ex}[1]{{\rm e}^{#1}} \def\ii{{\rm i}}
\def\rr{{\rm r}} \def\rs{{\rm s}}\def\rv{{\rm v}}
\def\ri{{\rm i}}\def\rj{{\rm j}}
\newcommand{\lrbrk}[1]{\left(#1\right)}
\newcommand{\sfrac}[2]{{\textstyle\frac{#1}{#2}}}

\def\Li{{\rm Li}_2}

\font\mybb=msbm10 at 12pt
\def\bb#1{\hbox{\mybb#1}}

\font\myBB=msbm10 at 18pt
\def\BB#1{\hbox{\myBB#1}}

\begin{document}

\begin{flushright}
QMUL-PH-09-14
\end{flushright}

\vspace{20pt}

\begin{center}

{\Large \bf Proof of the  Dual Conformal Anomaly of    }
\\
\vspace{0.3cm}
{\Large \bf One-Loop  Amplitudes in  $\cN=4$ SYM   }
\vspace{11pt}
\vspace{32pt}

{\mbox {\bf Andreas Brandhuber, Paul Heslop and Gabriele Travaglini}}%
\footnote{
{\sffamily \{\tt a.brandhuber, p.j.heslop, g.travaglini\}@qmul.ac.uk }}

{\em Centre for Research in String Theory\\
Department of Physics\\
Queen Mary, University of London\\
Mile End Road, London, E1 4NS\\
United Kingdom
 }

\vspace{30pt} {\bf Abstract}

\end{center}

\noindent
We provide two derivations of the one-loop dual conformal anomaly of generic $n$-point  superamplitudes in maximally supersymmetric Yang-Mills theory. Our proofs are based on simple applications of unitarity, and the known analytic properties of the amplitudes.

\setcounter{page}{0}
\thispagestyle{empty}
\newpage


\section{Introduction   }
\setcounter{footnote}{0}

A novel  symmetry of the planar S-matrix of $\cN=4$ supersymmetric
Yang-Mills (SYM) --  dual superconformal symmetry -- has been
introduced in \cite{dhks}. There, it was  conjectured  to be an exact
symmetry at tree level, but broken by quantum corrections, and an
expression for the anomaly associated to the dual conformal generators
was proposed.    
A confirmation of the conjecture was presented shortly after in \cite{bhtrec}, where it was demonstrated that the tree-level 
S-matrix of $\cN=4$ SYM transforms covariantly under the dual superconformal group. Furthermore, it was shown in the same paper that the supercoefficients, which appear in the expansion of planar, one-loop amplitudes in a basis of box functions, transform covariantly under the symmetry, {\it i.e.}~exactly in the same way as superamplitudes. 

Dual conformal symmetry was first observed in the context of the duality between MHV scattering amplitudes and Wilson loops \cite{am,dks,bht}. Strong indications of this duality were discovered in string theory in \cite{am}, where  the calculation of scattering amplitudes at strong coupling was mapped to that of a Wilson loop with a particular polygonal contour which can be constructed by gluing together the null momenta of the scattered particles, following the order of the insertions of the string vertex operators on the worldsheet.  Quite surprisingly, several calculations in perturbative 
$\cN=4$ SYM, first at one \cite{dks,bht} and then at two loops \cite{dhks4,dhks5,dhksbum,dhks6}, 
showed that the same duality holds also at weak coupling, 
with  perfect agreement found between the perturbative Wilson loop and the 
MHV scattering amplitudes of the $\cN=4$ theory computed in  \cite{bddk,2l5pt,seven,abdk}. 
The perturbative Wilson loop/amplitude duality was recently studied in 
\cite{Anastasiou:2009kn}, where a
numerical calculation of Wilson loops at two loops for an arbitrary
number of particles was presented.

At strong coupling, the emergence of dual superconformal symmetry  was understood in \cite{bermal,tse} 
using a peculiar T-duality of the superstring theory on $AdS_5 \times S^5$, which combines 
bosonic \cite{am} and fermionic T-duality transformations. The combined
effect of these T-dualities maps the original string sigma model into a dual sigma
model identical to the original one. More importantly, the T-duality also exchanges the original with the dual superconformal symmetries.

At weak coupling, dual conformal symmetry emerged as the ordinary conformal symmetry of the Wilson loop, which acts in the conventional way on 't Hooft's region  momenta  $x_i$.  These are defined via the relations $p_{i, \a \dot{\a}} = (x_{i}- x_{i+1})_{\a \dot{\a}}$, where $i=1, \ldots , n$, $n$ is the number of scattered particles,  and the  identification  $x_{n+1} = x_1$ enforces momentum conservation. 
Dual conformal symmetry is broken by loop effects and the corresponding anomalous Ward identity for the Wilson loop was derived in \cite{dhks4,dhks5}. 
In particular, it was shown in \cite{dhks5} that the ABDK/BDS ansatz \cite{abdk,bds} for the all-loop MHV amplitudes in $\cN=4$ SYM is a solution to this anomalous Ward identity. 

The origin of the dual conformal anomaly at the quantum level can be traced to the presence of cusps in the polygonal contour of the Wilson loop. For a smooth contour
dual conformal transformations would be an exact symmetry, however the 
cusps give rise to short-distance singularities which need to be regularised, and, hence, generate an anomaly in the dual conformal transformations at the loop level. These ultraviolet divergences are mapped to the conventional infrared divergences of the scattering amplitudes \cite{ir1,ir2,ir3,ir4,ir5,ir6,ir7,ir8}, thus suggesting an intimate  link between infrared singularities and the dual conformal anomaly.

Inspired by the anomaly derived from the Wilson loop side and the duality with MHV amplitudes, 
dual conformal symmetry was extended in \cite{dhks} to dual superconformal symmetry, acting on superamplitudes \cite{Nair} defined in a dual on-shell superspace. 
Moreover, it was suggested that any superamplitude factorises naturally into the MHV superamplitude
and a dual superconformal invariant factor $\cR$ as $ \cA =  {\cA}_\mathrm{MHV} \,  \cR$. 
The MHV superamplitude factor completely encapsulates the anomaly,
which is therefore a universal quantity. 
This remarkable conjecture was checked at one loop in \cite{dhksgen}
for the next-to-MHV (NMHV) superamplitudes up to nine particles. Very
recently in \cite{Brandhuber:2009xz,Elvang:2009ya} dual conformal
covariance was proved for one-loop NMHV superamplitudes with an arbitrary number of external particles. 

The goal of this paper is to prove the dual conformal anomaly for
generic (non-MHV) one-loop superamplitudes in the $\cN=4$ theory. In
order to do so, we will build on the results of
\cite{Brandhuber:2009xz}, where the most generic expression for the
dual conformal anomaly of all $\cN=4$ superamplitudes was derived
using only the result that the superamplitude can be expanded in terms
of box functions \cite{bddk}.  
The result of that calculation, reviewed in Section 2,  was found to be the sum of two terms. The first one is precisely the one-loop anomaly conjectured in \cite{dhks}.   
Therefore, the additional term must vanish if the conjecture of \cite{dhks} is correct. This indeed happens for all MHV and NMHV superamplitudes, as was proved in \cite{Brandhuber:2009xz} by using the explicit forms of these amplitudes derived in  \cite{bddk} and in \cite{dhksgen}. 
Moreover,  a new set of equations for the one-loop supercoefficients of a generic non-MHV amplitude were derived  in \cite{Brandhuber:2009xz}  by assuming the vanishing of this additional term. 
In this paper we will prove that this term does indeed vanish for generic superamplitudes, thus providing a proof of the dual conformal anomaly conjectured in \cite{dhks} at the one-loop level and, consequently, of the conformal equations presented in \cite{Brandhuber:2009xz}.

As will be explained in Section 2,
this additional term is finite in four dimensions and can be written as a particular linear combination of two-mass triangle functions, which depend on multi-particle as well as two-particle invariants. On the other hand, the dual conformal anomaly of \cite{dhks} diverges as $1/\eps$ as $\eps \to 0$, and depends only on
two-particle invariants through one-mass triangles. 
Note that we work here in dimensional regularisation with $D = 4 - 2 \eps$.
The different analytical structures of the two terms in the anomaly suggest that it is sufficient to study the discontinuities of the anomaly in all possible kinematic channels in order to prove that the additional term in the anomaly is in fact absent. Importantly, these discontinuities can be expressed in terms of appropriate phase space integrals.
In Section 3 we will begin by calculating two-particle cuts of the
one-loop anomaly for a generic superamplitude which are associated to discontinuities in multi-particle channels.%
\footnote{This has been done in the recent paper \cite{Korchemsky:2009hm} for MHV amplitudes. } 
We will find  that for any superamplitude these multi-particle discontinuities give rise to finite phase space integrals multiplied by $\eps$.
Hence, all the  multi-particle discontinuities of the dual conformal variation of a superamplitude vanish in four dimensions. 
With this result we  can rule out any additional terms to the anomaly conjectured in 
\cite{dhks}. We emphasise that our proof is general and applies to
superamplitudes with arbitrary total helicity and an arbitrary number
of external particles.

We have mentioned earlier a potential link between the dual conformal anomaly and infrared divergences. 
In Section 4 we expose this connection further by considering two-particle cuts of the anomaly in two-particle channels. 
Unlike the multi-particle discontinuities discussed above, the
two-particle discontinuities 
of the anomaly are non-zero and finite in four dimensions. By
uplifting the cut to a full loop diagram, akin to a procedure
introduced in \cite{Kosower:1999xi} for the calculation of splitting
amplitudes, we calculate its leading infrared divergence, which in
this case is of the order $1/\eps$. This turns out to reproduce
precisely the anomaly of \cite{dhks}. 

Our treatment of the two-particle channels exposes the leading  $1/\eps^2$ infrared singularity in this uplifted one-loop integral 
(which is further multiplied by one power of $\eps$ from an anomalous Jacobian), and in principle could miss subleading 
$1 / \epsilon$ contributions to it; these, in turn, would lead to finite, unwanted  contributions to the anomaly.  
However, we will argue that, thanks to the  no-triangle and bubble
property of one-loop amplitudes in $\cN=4$ SYM \cite{bddk},  our
approximation  in fact captures all the infrared divergences of the
above mentioned loop integral.  This result, together with the absence
of discontinuities of the anomaly in multi-particle channels, will
provide us with a second (albeit intimately related) proof of the dual conformal anomaly conjectured in \cite{dhks}. 

This second proof has the virtue of making more manifest the
connection of the dual conformal anomaly of a generic scattering
amplitude to its  infrared divergences. We stress that  a crucial
ingredient of both proofs is the maximal  supersymmetry of the theory.
In the second proof, this enters directly through the no-triangle and
bubble property of the $\cN=4$ amplitudes \cite{bddk}; in the first
proof, it enters through the specific form of the most general
anomaly, derived in \cite{Brandhuber:2009xz}.\footnote{This form of
  the anomaly can
  itself be thought of as a non-trivial consequence of the no triangle
  and bubble 
  property.}

\section{Background}
In this section we will first describe the structure of  one-loop superamplitudes, and will then discuss the general form of the dual conformal anomaly. 

\subsection{One-loop superamplitudes}

Scattering amplitudes in  $\cN=4$ SYM  with a fixed number of particles and total helicity are naturally  
combined into  superamplitudes \cite{Nair}. 
These are defined in an on-shell superspace where to each particle $i$ one associates the 
momentum $p_i =\lambda_i \tilde{\lambda}_i$, as well as a fermionic variables $\eta^{A}_i$, where 
$A=1, \ldots , 4$ is an $SU(4)$ index. 
The superamplitude can then be expanded in powers of the $\eta^{A}_i$'s, and  each term of 
this expansion corresponds to a particular amplitude in 
$\cN=4 $ SYM with  a fixed total helicity 
$h_{\rm tot} = \sum_{i=1}^n h_i$.  A term containing $m_i$  powers of $\eta_i$  corresponds 
to a scattering process  where the $i^\mathrm{th}$ particle has helicity 
$h_i = 1 - m_i/2$.
For instance,  the MHV superamplitude is given by the following compact expression 
\beq
\cA_\mathrm{MHV} \ = \ i\,  (2\pi)^4 \, {\delta^{(4)} (P) \, \delta^{(8)} (\Lambda) \over \lan 12\ran \lan 23\ran \cdots \lan n1\ran}
\ , 
\eeq
where $P:= \sum_{i=1}^n \l_i \lt_i$ and $\Lambda:= \sum_{i=1}^n \eta_i \l_i$ are the total momentum and supermomentum, respectively.

\begin{figure}[ht]
\begin{center}
\scalebox{0.55}{\includegraphics{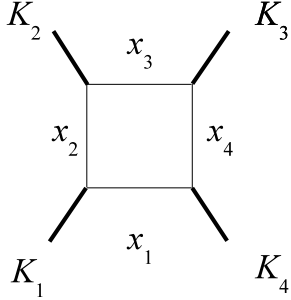}}
\end{center}
\caption{\it  A generic box function. 
$K_1, K_2, K_3$ and $K_4$  denote the  external momenta, and $x_1, x_2, x_3$ and $x_4$  the
corresponding region momenta, with $K_i = x_i - x_{i+1}$, $i=1, \ldots , 4$. }
\label{boxfunction}
\end{figure}

One-loop amplitudes in the maximally supersymmetric $\cN=4$  theory can be expanded in a known basis of  integrals
which contains only box functions%
\footnote{We use a collective index $i$ to denote the box function with external momenta 
$K_{1\ldots 4}$.}
 $F_i$, and no triangle or bubble functions \cite{bddk}. 
The functions $F_i$   are related to the 
scalar box integrals $I_i$ by a kinematic prefactor as follows. 
We call  $K_1, K_2, K_3$ and $K_4$  the external momenta at the four corners of a given box function,  
which are expressed as sums of momenta $p_i$ of external
particles. The momenta 
$K_{1\ldots4}$ can also be written  in terms of the region momenta 
$x_{1\ldots4}$, e.g. $K_1 = x_{12}$, where $x_{ij} := x_i-x_j$ (see Figure \ref{boxfunction}).
Then, up to a numerical constant, the relation between the $F$'s and the $I$'s  is
\begin{eqnarray}
\label{bpm}
I_i &= &-2\, \frac{F_i}{\sqrt{R_i}} \ , \nonumber \\
R_i&=& (x_{13}^2 x_{24}^2)^2 - 2 x_{13}^2 x_{24}^2  x_{12}^2  x_{34}^2  
- 2 x_{13}^2 x_{24}^2  x_{23}^2  x_{41}^2 +(x_{12}^2 x_{34}^2 - x_{23}^2 x_{41}^2)^2 \ .
\end{eqnarray}
Four-mass boxes are special from the point of view of the dual conformal symmetry, as they 
are infrared finite and  invariant under the symmetry.  
We can then  simplify the expression for $\sqrt{R}$ to 
\beq 
\label{Rcasopar}
\sqrt{R} \ \to \ x_{13}^2 x_{24}^2 - x_{23}^2 x_{41}^2 
\ , 
\eeq
valid for all box functions except four-mass ones in the case 
where either $x_{12}^2$ or $x_{34}^2$ vanish. Notice that,  under dual conformal  inversions, one has 
\be\label{rtrafo}
\sqrt{R_i} \ \to \  \frac{\sqrt{R_i}}{x_1^2 x_2^2 x_3^2 x_4^2} \, .
\ee
We  expand   a generic  $n $-point one-loop superamplitude $\mathcal{A}^{\mathrm{1-loop}}_n $ in terms of box functions \cite{bddk} as 
\begin{equation}
\label{exp3}
 {\mathcal{A}}^{\mathrm{1-loop}}_n \ =  \  
\sum_{\{i , j , k , l\}}  
{c}(i,j,k,l)  \,
F(i,j,k,l) \ ,
\end{equation}
where $i$, $j$, $k$, $l$, denote the four region momenta of the box function (as in Figure \ref{boxfunction}, with 
the labels $1$, $2$, $3$, $4$, replaced by $i$, $j$, $k$, $l$).

In \cite{bhtrec} it was shown that the supercoefficients ${c}(i,j,k,l)$ transform covariantly under the symmetry. 
In order to deal with quantities which are invariant under dual conformal transformations rather than covariant,
 it is convenient to  
redefine the dual conformal generator $K^{\mu}$  as \cite{dhp}  
\beq 
\label{khat}
K^\mu \ \to \ \hat{K}^\mu \ := \ K^{\mu} \, -\,  2\sum_{i=1}^n x^\mu_i
\ . 
\eeq
The covariance of the one-loop supercoefficients is then re-expressed as   
\beq
\label{spar}
\hat{K}^\mu{c}(i,j,k,l) =0
\ . 
\eeq

\subsection{The structure of the anomaly}

Dual conformal symmetry is violated at the quantum level by the presence of infrared divergences. In \cite{Brandhuber:2009xz}, using the expansion \eqref{exp3} of a generic superamplitude in a basis of boxes, together with the covariance of the one-loop supercoefficients \eqref{spar}, the dual conformal anomaly of an arbitrary superamplitude was written as
\beq
\label{2.9}
\hat{K}^\mu  {\mathcal{A}}^{\mathrm{1-loop}}  \ = \ 
\sum_{\{i , j , k , l\}}  
{c}(i,j,k,l)  \  K^\mu
F(i,j,k,l) \ . 
\eeq
After calculating the various box anomalies $K^\mu F(i,j,k,l)$,  \eqref{2.9} takes  the form \cite{Brandhuber:2009xz}
\beqa
\label{eq:8}
\hat{K}^{\mu}  {\mathcal{A}}^{\mathrm{1-loop}}_n
&=&4  \epsilon   \, \cA_{n}^\mathrm{tree} 
\sum_{i=1}^n
x_{i-1}^{\mu}\, x_{ii-2}^2\,J(x_{ii-2}^2) 
 \\
 &&\!\!\!\!\!-\, 2\epsilon   \sum_{i=1}^n \sum_{k=i+2}^{i+n-3} 
\cE(i,k)
\Big[  x_{i-1}^{\mu}\,
x_{ik}^2\, -\,  x_{i}^{\mu}\,
x_{i-1\,k}^2\Big] J(x_{i k}^2,x_{i-1\, k}^2)
\ , 
\nonumber
\eeqa
where 
\begin{align}
\label{Edef}
  \cE(i,k):= \sum_{j=k+1}^{i+n-2}
 {c}(i,k,j,i-1)-\sum_{j=i+1}^{k-1}
 {c}(i,j,k,i-1)\ ,
\end{align}
is a particular combination of supercoefficients. 
Furthermore, \eqref{Edef} is valid for $i<k$; if $i>k$, then
the variable $k$ appearing in the summation ranges of  \eqref{Edef} has
to be replaced by $k+n$. We have also introduced one-mass and two-mass triangle functions, defined as 
\beqa
J(a) &:= & \ {r_\Gamma\over \epsilon^2 }
 (-a)^{-\epsilon -1} \ ,  
\\
J(a, b) &:= & {r_\Gamma\over \epsilon^2 }
 { (-a)^{-\epsilon} - (-b)^{-\epsilon}\over (-a) - (-b)   } 
\  , 
\end{eqnarray}
respectively, where $r_\Gamma := \Gamma (1 + \e) \Gamma^2 ( 1 - \e )  / \Gamma (1 - 2 \e) $.

Equation \eqref{eq:8} gives the most general expression for the anomaly of a  one-loop superamplitude in $\cN=4$ SYM with an arbitrary total helicity. 
In order to set the scene for our proof, let us now highlight the main characteristics of \eqref{eq:8}.

To begin with, the first term of  \eqref{eq:8} precisely matches the anomaly conjectured in \cite{dhks}. Furthermore, it contains only one-mass triangles whose arguments are two-particle invariants. 
These triangles, multiplied by $\e$, give rise to terms which diverge as $1/\eps$ as $\eps\to 0$. 

On the other hand, the second line of \eqref{eq:8}  contains two-mass triangles 
whose arguments can be two- or multi-particle invariants. In general, there is no two-mass triangle that has only two-particle invariants, except at five points, where amplitudes are only MHV or anti-MHV. This specific case has already been addressed explicitly in \cite{dhks,Brandhuber:2009xz} where it was shown that the anomaly of \cite{dhks} is correctly reproduced. 

The presence of multi-particle invariants in the second line of \eqref{eq:8} is its key signature, and in the next section  we will use the analyticity properties of this expression to prove that, in fact, this term identically vanishes. As a byproduct, this implies the conformal equations 
\beq
\cE(i,k) \ = \ 0 \ ,  \qquad  i=1,
\ldots, n\ ,\ \ k=i+2, \ldots , i+n-3 \ , 
\label{gennconfeq2}
\eeq
relating box coefficients, where $\cE(i,k) $ are given in \eqref{Edef}.
These relations were conjectured in \cite{Brandhuber:2009xz} to hold for any superamplitude, 
and checked explicitly for the infinite sequences of MHV and NMHV
superamplitudes. They can be solved to give expressions for all
one-mass, two-mass easy 
and half of the two-mass hard box coefficients in terms of the
remaining box coefficients.

\section{The first proof} 

We perform the proof of the one loop dual conformal anomaly in two steps:
 
  {\bf 1.} We will calculate the discontinuities of the anomaly using
  conventional unitarity \cite{fusing}, and prove that for
  multi-particle channels, the result for the discontinuity is
  given by $\e$ times an integral which is finite in four
  dimensions. The result for such a discontinuity therefore vanishes
  in four dimensions.
  
  {\bf 2.} We will calculate the discontinuity of the anomaly in a
  multi-particle channel directly from \eqref{eq:8}, and impose that
  this vanishes.    This  precisely implies the dual conformal equations  \eqref{gennconfeq2}
and therefore proves the form of the anomaly
conjectured in \cite{dhks} for all one-loop superamplitudes in the $\cal N$=4 theory, 
\beq
\label{eq:trueanom}
\hat{K}^{\mu}  {\mathcal{A}}^{\mathrm{1-loop}}_n
\ = \ 4  \epsilon   \, \cA_{n}^\mathrm{tree} 
\sum_{i=1}^n
x_{i-1}^{\mu}\, x_{ii-2}^2\,J(x_{ii-2}^2) 
\ . 
\eeq

We now proceed directly to the proof.

\begin{figure}[ht]
\begin{center}
\scalebox{0.55}{\includegraphics{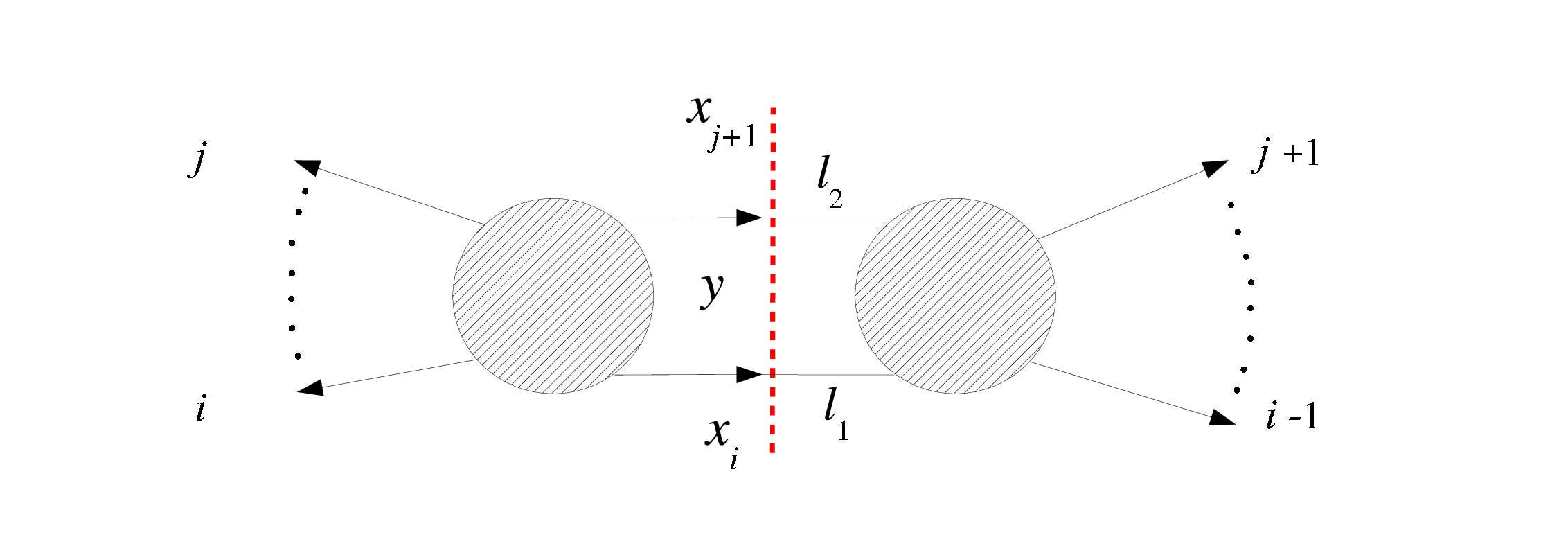}}
\end{center}
\caption{\it  A cut diagram reproducing the discontinuity of the anomaly for a generic superamplitude in a kinematic  channel $x_{i\,j+1}^2$. When $x_{i\,j+1}^2$ is a multi-particle invariant, the phase space integral corresponding to this cut diagram is finite, and vanishes in four dimensions due to the factor of $D-4$  on the right hand side of \eqref{anomalycut}.}
\label{multipartcut}
\end{figure}

{\bf 1. } Consider the discontinuity of the superamplitude in a
certain multi-particle channel $P_L^2$. We wish to show that this is
conformally invariant (this is not true for the two-particle channel -- 
such cuts will be considered in the following  section).
 The corresponding cut diagram is represented in Figure
 \ref{multipartcut}, and is  expressed by the following phase space
 integral:  
\beq
\label{supercut}
\int\! d\mu_{i, \ldots , j} 
\ 
\cA_L (l_2, l_1, i, \ldots , j) \, \cA_R (-l_1, -l_2, j+1, \ldots , i-1) 
\ . 
\eeq
The integration measure is  defined as 
\beq
\label{muu}
d \mu_{i, \ldots , j}  \ := \  d\mathrm{LIPS}(l_2, l_1; P_L)  \ d^4 \eta_{l_1} d^4 \eta_{l_2} \delta^{(8)} ( \eta_{l_1} \l_{l_1} +  \eta_{l_2} \l_{l_2} + \Lambda_L ) 
\ , 
\eeq
where 
\beq
d\mathrm{LIPS}(l_2, l_1; P_L)  = \, d^Dl_1 d^D l_2 \, \delta^{(+)} (l_1^2) \, \delta^{(+)} (l_2^2)\ \delta^{(D)} (l_1 + l_2 + P_L) 
\ , 
\eeq
is the phase space measure, and 
$d^4 \eta_{l_1} d^4 \eta_{l_2} \delta^{(8)} ( \eta_{l_1} \l_{l_1} +  \eta_{l_2} \l_{l_2} + \Lambda_L ) $ is the fermionic integration measure. 
We have defined 
\beqa
P_L  \ := \  \sum_{k=i}^{j} \lambda_k  \lt_k \ , \qquad 
\Lambda_L  \ := \  \sum_{k=i}^{j} \eta_k \lambda_k  \ , 
\eeqa
to be the total momenta and supermomenta flowing out of  the left hand side of the cut diagram. In \eqref{supercut} we are omitting overall delta functions imposing momentum and supermomentum conservation.

In the appendix we compute the dual conformal transformation
of the discontinuity 
in the $x_{i\,j+1}^2$ channel, with the result
\beq
\label{anomalycut}
\mathrm{disc}_{x_{i\,j+1}^2}\big[\hat K^\mu\cA^{\rm 1-loop}_n \big] = 2(4-D) \, \int\!\!d^D y \, \delta^{(+)}\big ((y-x_i)^2\big) \, \delta^{(+)} \big((x_{j+1}-y)^2\big) 
\ \Big[ y^\mu \,  \langle l_1 l_2
\rangle^4 \cA_L \cA_R \Big] \, .
\eeq

In order to understand whether \eqref{anomalycut} leads to a contribution to the anomaly, we analyse the singularities of the integral in that equation. To this end,  we first consider the phase space integral giving the discontinuity of the superamplitude in the same channel 
$x_{i\,j+1}^2$. This is given by 
\beq
\label{discontpij}
\mathrm{disc}_{x_{i\,j+1}^2}\cA^{\rm 1-loop}_n \, = \, \int\!\!d^D y \, \delta^{(+)}\big ((y-x_i)^2\big) \, \delta^{(+)} \big((x_{j+1}-y)^2\big) 
\ \Big[  \langle l_1 l_2
\rangle^4 \cA_L \cA_R \Big] \ .
\eeq

As is well known, there is a crucial distinction in the infrared properties of \eqref{discontpij} 
between the cases when the channel is a multi-particle or a two-particle one. 
When $x_{i\,j+1}^2$ is a multi-particle channel, the integral appearing on the right hand side of \eqref{discontpij} 
is free of infrared divergences  and hence can be calculated in four dimensions, 
see \cite{bbkr} and \cite{Kosower:1999xi} for a discussion of this point.%
\footnote{For the sake of this first proof, we are only interested in multi-particle channels, as discussed above. 
We will later on discuss the two-particle channel discontinuities, to show how the anomaly arises precisely from such singular channels. } 
This is of course in agreement with the general expression of the infrared divergences of one-loop amplitudes 
in  $\cN=4$ SYM, given by  \cite{Kunszt:1994mc}
\beq
\label{ircr} 
\mathcal{A}^{\mathrm{1-loop}}_n |_{\mathrm{IR}} \ = \ - r_{\Gamma} \, 
\mathcal{A}^{\mathrm{tree}}_n \sum_{i=1}^n 
\frac{(-x_{i i+2}^2)^{-\epsilon}}{\epsilon^2} \ , 
\end{equation}
which only contain two-particle invariants formed with adjacent momenta (but no multi-particle invariant).

We now make the observation that the integral we are really interested in, namely that on the right hand side of 
\eqref{anomalycut} is very similar to the integral appearing in
\eqref{discontpij}.   More precisely, \eqref{anomalycut} contains an
extra power of $2(D-4)$ and a $y^\mu$ in the integrand compared to
\eqref{discontpij}. 
The presence in  \eqref{anomalycut} of  $y= x_{j+1} - l_1 $ by itself  cannot lead to any infrared
singularity (the term containing $l_1$ will only give rise to terms which are better behaved in the infrared). 
Since \eqref{discontpij} is infrared finite, we conclude that  the presence of a factor of $D-4$ multiplying 
the discontinuity of the anomaly \eqref{anomalycut} will make the result vanish. 
Hence,  for  a generic multi-particle channel $x_{i\,j+1}^2$, 
\beq
\label{zero}
\mathrm{disc}_{x_{i\,j+1}^2} \big[\hat K^\mu\cA^{\rm 1-loop}_n \big]\,  = \, 0 \ , \quad j \neq i+1 
\ . 
\eeq
This concludes the first part of the proof. Notice that \eqref{zero}  is in
agreement with \cite{Korchemsky:2009hm}, where it was observed  
that the discontinuities of the one-loop MHV amplitude in multi-particle
channels, calculated in \cite{bbkr},  are dual conformal invariant. Our result \eqref{zero} is however completely general, 
in that it applies to all one-loop amplitudes, including non-MHV.

{\bf 2.} We now wish to use the absence of conformal anomalies in the
multi-particle cut \eqref{zero}  to  constrain the expression  \eqref{eq:8}.  More precisely, we will use the fact that in each of the $n(n-5)/2$ multi-particle channels the discontinuity of the anomaly vanishes in order to prove the conformal equations \eqref{gennconfeq2}, and hence the form of the conformal anomaly for generic amplitudes. 

To this end, we focus on the terms on the right hand side of \eqref{eq:8} which have a discontinuity in a certain multi-particle channel $x_{ik}^2$. There are four such terms: 
\beqa
\hat K^\mu\cA^{\rm 1-loop}_n  &\ni &\!\!\!\!-2\eps\, 
\cE(i,k) \, 
\Big[  x_{i-1}^{\mu}\, x_{ik}^2\, -\,  x_{i}^{\mu}\,
x_{i-1\,k}^2\Big] J(x_{i k}^2,x_{i-1\, k}^2)\nonumber \\ \nonumber 
& - & 
2\eps\, \cE(i+1,k) \, 
\Big[  x_{i}^{\mu}\, x_{i+1\, k}^2\, -\,  x_{i+1}^{\mu}\,
x_{ik}^2\Big] J(x_{i+1\,  k}^2,x_{i k}^2)
\nonumber \\ 
&-&2\eps\,  \cE(k,i) \, 
\Big[  x_{k-1}^{\mu}\, x_{ik}^2\, -\,  x_{k}^{\mu}\,
x_{k-1\,i}^2\Big] J(x_{i k}^2,x_{k-1\, i}^2)\\ \nonumber 
& - & 
2\eps\, \cE(k+1,i) \, 
\Big[  x_{k}^{\mu}\, x_{k+1\, i}^2\, -\,  x_{k+1}^{\mu}\,
x_{ik}^2\Big] J(x_{k+1\,  i}^2,x_{ik}^2)\ . 
\eeqa 
The last two lines are obtained from the first two by simply exchanging $i$ with $k$. 
The discontinuity of a triangle function is given by
\beq
\mathrm{disc}_b \,  \big[ \eps \, J(a, b)\big]\ = \ {2\pi i   \over b-a} \, + \, \cO(\eps)\ , 
\eeq
therefore 
\beqa
&&\hspace{-1cm}\mathrm{disc}_{x_{ik}^2 }\big[K^\mu\cA^{\rm 1-loop}_n \big]\nonumber \\ 
&=& 
2 \pi i \Big[ \cE(i,k) \, 
{ x_{i-1}^{\mu}\, x_{ik}^2\, -\,  x_{i}^{\mu}\,
x_{i-1\,k}^2\over x_{i k}^2- x_{i-1\, k}^2} +  
 \cE(i+1,k) \, 
{ x_{i}^{\mu}\, x_{i+1\, k}^2\, -\,  x_{i+1}^{\mu}\,
x_{ik}^2\over 
x_{i k}^2-x_{i+1\,  k}^2} 
\nonumber \\
&& \phantom{2 \pi i \Big[} \cE(k,i) \, 
{x_{k-1}^{\mu}\, x_{ik}^2\, -\,  x_{k}^{\mu}\,
x_{k-1\,i}^2\over  x_{i k}^2- x_{k-1\, i}^2 }
 \ + \ \cE(k+1,i) \, 
{ x_{k}^{\mu}\, x_{k+1\, i}^2\, -\,  x_{k+1}^{\mu}\,
x_{ik}^2\over x_{ik}^2-  x_{k+1\,  i}^2}\Big] 
\nonumber \\
&=&0
\, , 
\label{vanishdiscont}
\eeqa 
where in the last step we have used \eqref{zero}. 

Equation \eqref{vanishdiscont} is a vector equation, 
which we can rewrite as 
\beq 
\cE (i, k) v_{ik}^\mu + \cE (i+1, k) v_{i+1k}^\mu +  \cE (k, i) v_{ki}^\mu + \cE (k+1, i) v_{k+1i}^\mu\ = \ 0
\ , \label{Ev}
\eeq
where  
\beq
v_{ik}^\mu \, :=\,  
\frac{ x_{i-1}^{\mu}\, x_{ik}^2\, -\,  x_{i}^{\mu}\, x_{i-1\,k}^2}{
x_{ik}^2 - x_{i-1 \, k}^2}
\ , 
\eeq
and we remark that the four vectors $v_{ik}^\mu$, $v_{i-1k}^\mu$,  $v_{ki}^\mu$ and $v_{k-1i}^\mu$ are in general linearly independent in four dimensions. Hence we conclude that the coefficients $\cE (i, k)$, $\cE (k,i)$, $ \cE (i+1, k)$, $ \cE (k+1, i) $ must vanish independently. 

This proves the conformal equations \eqref{gennconfeq2},  and therefore we conclude that the 
dual conformal anomaly for an arbitrary (non-MHV) amplitude is given by \eqref{eq:trueanom}. 

This completes our proof.   We conclude this section with a couple of additional comments.

First, we have managed to prove $n(n-4)$ conformal
equations $\cE(i,k)$=0 in 
\eqref{gennconfeq2}  from
considering just $n(n-5)/2$ multi-particle cuts of the
amplitude. 
We can do this since
each multiparticle cut anomaly is a vectorial equation and hence gives four
independent conditions, thus we really obtain  $2n(n-5)$ (dependent)
conditions. These are precisely the conditions \eqref{Ev}. 
Each multiparticle cut $x_{ik}^2$ leads to the condition $\cE(i,k) =0 $ and
$\cE(k,i)=0$. This leaves only the `boundary case' $\cE(i,i+2)=0$
potentially unaccounted for. Fortunately \eqref{Ev} also gives
$\cE(i+1,k)=0$ which 
for $k=i+3$ gives us precisely this boundary case with the (arbitrary)
label $i$ shifted to $i+1$.

Second, we mention an important point which could have affected our proof. 
It has been recently pointed out in \cite{new,Korchemsky:2009hm} 
that some of the dual superconformal generators 
do not precisely annihilate the superamplitude but leave behind 
a delta-function supported contribution of the type of an  holomorphic anomaly \cite{holo}. 
It is important for our proof that there is no holomorphic anomaly for the special conformal generator ${K}^\mu$ 
we are interested in. Indeed, had  ${K}^\mu $ acting on the tree-level amplitudes $\cA_L$ or $\cA_R$ in \eqref{discontpij} produced delta-function contributions, the phase space integration in \eqref{discontpij} would be localised, and new, unwanted contributions to the dual conformal anomaly would be generated.

Fortunately, the absence of holomorphic anomaly contributions to the dual special conformal generator 
$K^\mu$ has been shown for all tree-level  amplitudes in \cite{new}, and specifically for MHV superamplitudes and six-point NMHV tree-level superamplitudes  in  \cite{Korchemsky:2009hm}. 
We recall that the holomorphic anomaly arises from 
\beq
{\partial \over \partial \lt^{\dot\alpha}} {1\over \lan \l \m\ran } \, = \, 
2 \pi \tilde{\mu}_{\dot\alpha} \, \delta ( \lan \l \m \ran ) \, \delta ( [\lt \tilde\mu]) 
 \ . 
\eeq
In order to get a contribution from holomorphic anomalies which would lead to a localisation of the phase space integral in \eqref{discontpij}, one should then identify all possible physical singularities in the scattering amplitude of the type
$1 / \lan l k  \ran$, where $l$ is any one of the cut loop momenta, and $k$ one of the legs which are adjacent to it.%
\footnote{We consider colour-ordered amplitudes, hence such singularities can only involve  particles 
that are adjacent in colour space.}  
At tree level, such singularities arise only from  collinear kinematics, 
and in \cite{Korchemsky:2009hm} it has been shown that in the action of 
$K_{\alpha \dot\alpha}$ on  $1 /  \lan i  i +1 \ran$, the holomorphic anomaly contributions cancel. 
Therefore, the special conformal generator is not affected by holomorphic anomalies.%
\footnote{Unlike the dual special conformal generator, the dual
  supersymmetry generator $\bar{Q}$ does suffer from a holomorphic anomaly \cite{Korchemsky:2009hm}.}

\section{Unearthing the anomaly with two-particle cuts}

To complete our discussion, we now address the two-particle channel cuts of
the conformal variation of the amplitudes in more detail.
This will reveal the close relation between infrared divergences of
the amplitude
and the dual conformal anomaly.
\begin{figure}[ht]
\begin{center}
\scalebox{0.55}{\includegraphics{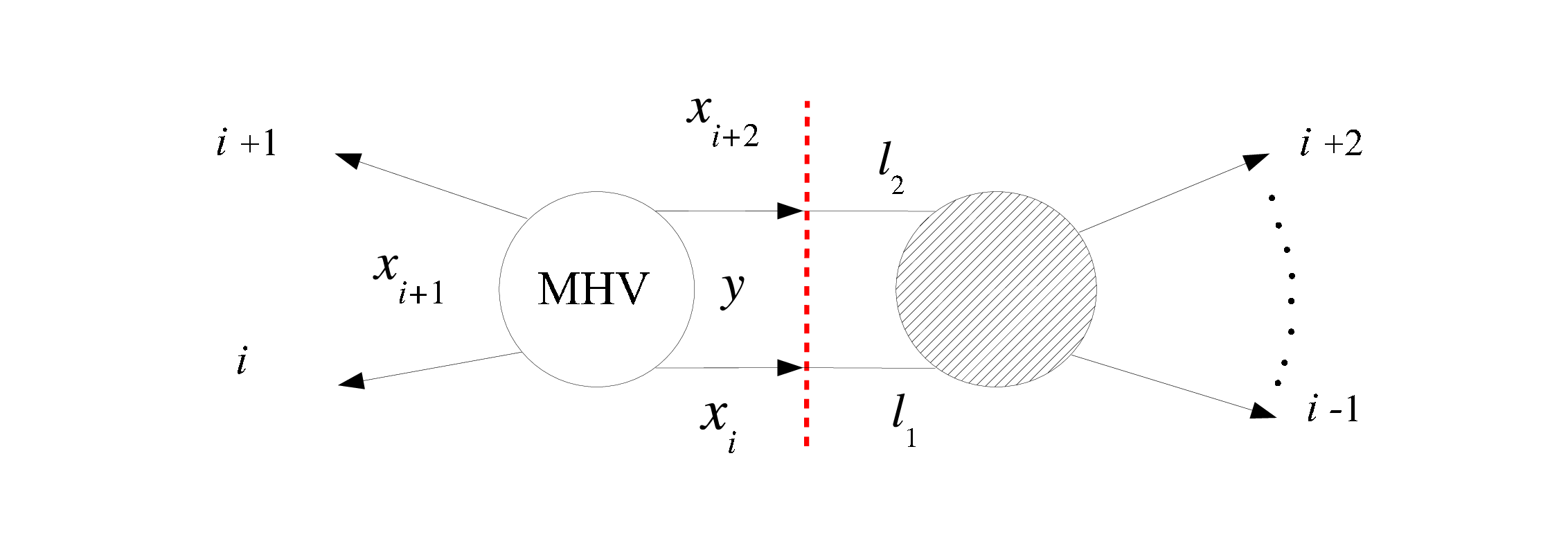}}
\end{center}
\caption{\it The cut diagram reproducing the discontinuity of the
anomaly for a generic superamplitude in the two-particle channel
$s_{i i+1}=x^2_{ii+2}$.  In this case the superamplitude on the left
hand side has four particles and, hence, must be a MHV superamplitude.
This diagram has an infrared divergence arising from the region of
integration where  $l_1\sim -p_i$ and $l_2\sim -p_{i+1}$. In this
region we also have that $y\sim x_{i+1}$ where $y$ is the region
momentum between the two cut legs.}
\label{twopartcut}
\end{figure}
For concreteness let us consider the two-particle channel cut of a generic
one-loop amplitude represented in Figure \ref{twopartcut}. This
 cut gives rise to the following phase space integral, 
\beq
\label{supercut2}
\mathrm{disc}_{x_{i i+2}^2} \cA^{\rm 1-loop}_n \ = \  \int\! d\mu_{i i+1}
\
\cA_{\rm MHV} ( i, i+1, l_2, l_1 ) \, \cA_R (-l_1, -l_2, i+2, \ldots , i-1)
\ ,
\eeq
while the corresponding cut diagram of the dual conformal anomaly
is given by
\beq
\label{anomalycut2p}
\mathrm{disc}_{x_{i\,i+2}^2}
\big[K^\mu\cA^{\rm 1-loop}_n \big] =
4\eps \, \int\!d\mu_{ii+1}
\Big[ y^\mu
\cA_\mathrm{MHV} ( i, i+1, l_2, l_1 ) \, \cA_R (-l_1, -l_2, i+2,
\ldots , i-1) \Big] \ ,
\eeq
where the integration measure appearing in both expressions is defined
in \eqref{muu}.

The phase space integrals appearing in \eqref{supercut2} and
\eqref{anomalycut2p} are both infrared divergent. The divergence
arises from a region where $l_1$ becomes collinear to $p_i$ and $l_2$
becomes collinear to $p_{i+1}$ at the same time \cite{Kosower:1999xi}.
This occurrence of simultaneous collinear singularities is special to
two-particle cuts and is necessary to produce the expected infrared
divergences.
For generic kinematics, {\it i.e.} if $p_i$ and $p_{i+1}$ are not
collinear, momentum conservation of the four-point MHV amplitude
implies that the singular region of the momentum integral is confined
to
\beq\label{singregion}
l_1 \, \to -p_i, \qquad l_2 \, \to \, -p_{i+1}
\ .
\eeq
In this singular region of loop momentum space, the region momentum $y$
localises on the region momentum $x_{i+1}$, which will be of
importance in the following.

\begin{figure}[ht]
\begin{center}
\scalebox{0.55}{\includegraphics{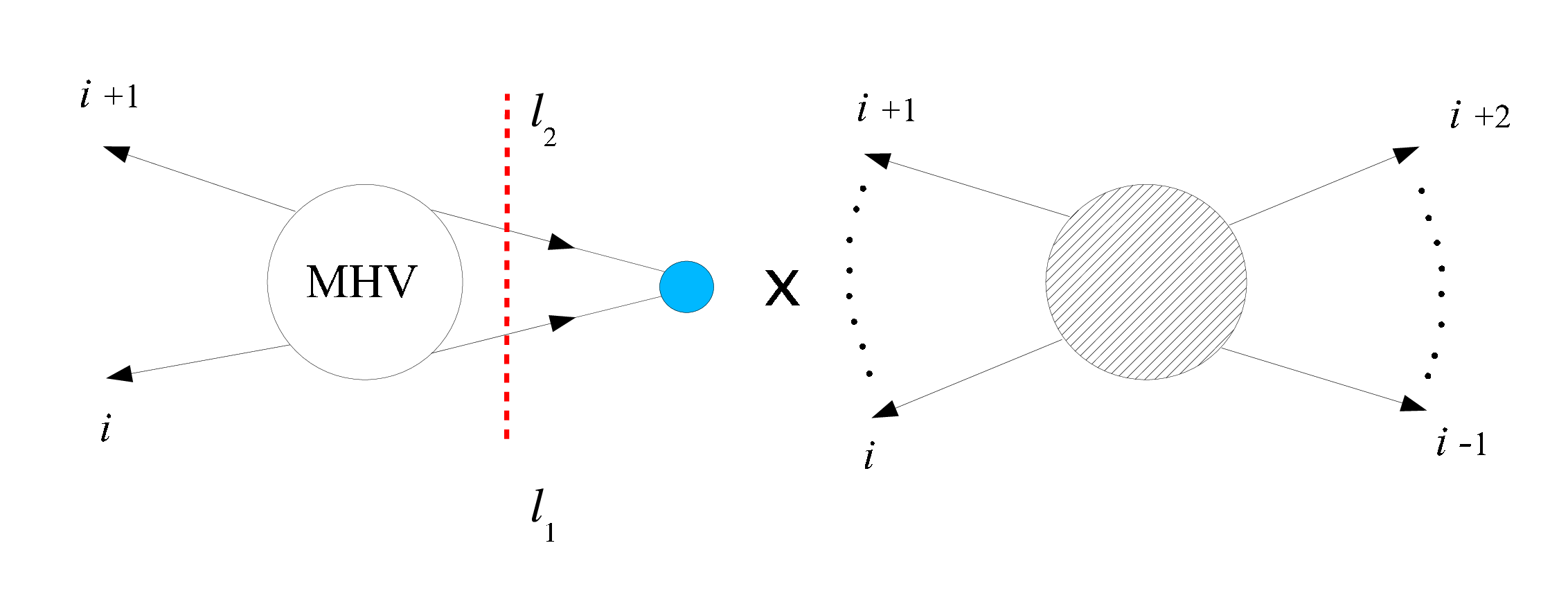}}
\end{center}
\caption{\it Graphic representation of the factorised structure of 
the two-particle channel discontinuities of the amplitude and of the anomaly in 
\eqref{ampcuttri} and \eqref{anomcuttri}. }
\label{efftwopartcut}
\end{figure}

In the following we will show that  the
exact one-loop anomaly and the infrared divergent part of the amplitude can
be extracted by using an approximation which focuses exactly on this
peculiar loop momentum region.
A couple of explanations are in order here:
\begin{itemize}
\item We observe that the leading infrared singularity of both
integrals is correctly captured by replacing $-l_1$ and $-l_2$ in the
tree-level superamplitude $\cA_{R}$ by
$p_i$ and $p_{i+1}$, respectively. Importantly, this
allows us to pull $\cA_{R}$ out in front of the integrals
\eqref{supercut2} and \eqref{anomalycut2p}. Note that this tree-level
superamplitude has the same total
helicity as the one-loop amplitude under consideration.
Furthermore, in the integrand of \eqref{anomalycut2p} we can replace
$y$ with $x_{i+1}$ and factor it out of the integral as well.
This factorisation property of the leading infrared singularity is
illustrated in Figure \ref{efftwopartcut}.
\item We know from \eqref{zero} in the previous section that the
anomaly does not have discontinuities in multi-particle channels and,
hence, it can only depend on
two-particle invariants.
\item
Our approximation of the cut integrals \eqref{supercut2}  and
\eqref{anomalycut2p} removes all dependence on momentum invariants other
than the
two-particle invariant $x_{ii+2}^2$. This allows us
to directly uplift the cut integrals to full loop integrals, 
{\it i.e.} replace the two on-shell $\delta$-functions in the cut by  propagators, very much as was done in \cite{Kosower:1999xi}
for splitting amplitudes.
\item
The remaining issue is to rule out any subleading infrared terms that our
approximation might
miss. For the amplitudes, this is easily settled by recalling that in
$\mathcal{N}=4$ SYM
all one-loop amplitudes are linear combinations of box functions and
that all potentially infrared divergent terms are expressed in terms of
one-mass triangle functions with a two-particle invariant argument, 
which behave as $(-x_{i i+2}^2)^{-\epsilon}/\eps^2$. 
Indeed, after uplifting 
the cut  integral, one obtains
one-mass triangle integrals with the expected coefficient. If we now apply
dual special conformal transformation on our integrand, one power of
the loop momentum appears in the numerator. Uplifting then gives rise
to  linear box functions, which can be reduced
to scalar boxes and triangles. 
The result consists of finite
terms and terms that are of the form $(-s)^{-\epsilon}/\eps^2$ or
$(-t)^{-\epsilon}/\epsilon^2$ for two- (multi-)particle invariants $s$
 $(t)$. Crucially, there are no subleading $1/\eps$ terms, {\it i.e.}  bubble functions.
Since the dual conformal transformation comes with a factor of $\eps$,
only one-mass triangles with two-particle invariants survive. 
So in essence the no-bubble and no-triangle property of one-loop amplitudes
in $\mathcal{N}=4$ SYM ensures that our procedure is
valid.%
\footnote{For the precise form of the reduced linear boxes see
  \eqref{eq:8}. However in this section we wish to give general
  arguments without referring to this formula for the generic anomaly. }
\end{itemize}

After these comments we proceed now to the explicit evaluation of the
two-particle cut integrals \eqref{supercut2} and
\eqref{anomalycut2p}. 
For the infrared divergent part of the amplitude we find
\begin{align}
\nonumber
 \left.\Big[ \cA^{\rm 1-loop}_{n}
\Big]_{x_{i i+2}^2-\mathrm{cut}} \right|_{\mathrm{IR}}
= &  \ \cA_n^\mathrm{tree}
 \int\!\!{d^D y\over (2\pi)^D} \, \ \Big[\, { \langle l_1  l_2
\rangle^4  \, \cA_\mathrm{MHV} ( i, i+1, l_2, l_1 ) \over
(x_i-y)^2(x_{i+2}-y)^2} 
 \Big]_{x_{i i+2}^2-\mathrm{cut}}  \ ,
\\ 
= & \ \cA_n^\mathrm{tree}
 \, \int\!\!{d^D y\over (2\pi)^D} \, \ \Big[\, {  \,  x^2_{i\,i+2}
  \over (x_i-y)^2(x_{i+1}-y)^2 (x_{i+2} -y)^2 }
 \Big]_{x_{i i+2}^2-\mathrm{cut}}  \ ,
 \label{ampcuttri}
\end{align}
where we have performed some spinor algebra and used
\eqref{singregion} to simplify the 
numerator of the integrand to obtain the second line.
We arrive at an expression that is
proportional to a scalar one-mass triangle.
Similarly, for the dual conformal anomaly of the amplitude we obtain
\begin{align}
\nonumber
\big[K^\mu\cA^{\rm 1-loop}_n \big]_{{x_{i i+2}^2}-\mathrm{cut}}  = &
4\eps  \,   x_{i+1}^\mu  \cA_n^\mathrm{tree} 
\, \int\!\!{d^D y\over (2\pi)^D} \ \Big[\, { \langle l_1  l_2
\rangle^4  \, \cA_\mathrm{MHV} ( i, i+1, l_2, l_1 ) \over
(x_{i+2}-y)^2 (x_i - y)^2}
 \Big]_{x_{i i+2}^2-\mathrm{cut}}\\ 
=& 4\eps  \,   x_{i+1}^\mu \cA_n^\mathrm{tree}
 \, \int\!\!{d^D y\over (2\pi)^D} \, \ \Big[\, {  \,  x^2_{i\,i+2}
  \over (x_i-y)^2(x_{i+1}-y)^2 (x_{i+2} -y)^2 }
 \Big]_{x_{i i+2}^2-\mathrm{cut}}.
 \label{anomcuttri}
\end{align}
We notice that these two relations establish a link between the
universal infrared behaviour of the amplitudes and the dual conformal
anomaly, since these imply that
\beqa
\nonumber
\label{anomalycut2plimreloaded}
\big[K^\mu\cA^{\rm 1-loop}_n \big]_{{x_{i i+2}^2}-\mathrm{cut}}  &
\!\!\!\!\!= \!\!&  4\eps\, x^\mu_{i+1}   \ \left.\Big[ \cA^{\rm
1-loop}_{n}
\Big]_{x_{i i+2}^2-\mathrm{cut}} \right|_{\mathrm{IR}}\ .
\eeqa
The expression \eqref{ampcuttri} can be freely uplifted and summed
over all two-particle channels (recalling that all multi-particle cuts
vanish) to give us directly the well known expression for
the infrared divergent part of the amplitude \eqref{ircr}. Similarly,
\eqref{anomcuttri} 
gives directly 
\beqa
K^\mu\cA^{\rm 1-loop}_n  &= &
4\eps \, \cA_n^\mathrm{tree} \sum_{i=1}^n x_{i+1}^\mu  \,
x_{ii+2}^2\,J(x_{ii+2}^2) \ .
\label{ultima}
\eeqa
The right hand side of \eqref{ultima} is nothing but the dual
conformal anomaly.
We have thus managed to link this anomaly to infrared-divergent
two-particle channel cut diagrams. This provides a derivation of the form of
the anomaly 
which exposes in a very direct
manner the exact coefficient of the anomaly.

\vspace{0.7cm}
\section*{Acknowledgements}

It is a pleasure to thank Bill Spence for stimulating  discussions. 
This work was supported by the STFC under a Rolling Grant  ST/G000565/1.
The work of PH is supported by an EPSRC Standard Research Grant EP/C544250/1.
GT is supported by an EPSRC Advanced Research Fellowship EP/C544242/1
and by an EPSRC Standard Research Grant EP/C544250/1.

\appendix

\section{Conformal transformation of a cut diagram}

We now consider the conformal variation of the discontinuity integral. This is given by the following expression, 
\beq
\label{superconf}
\int\! d\mu'_{i, \ldots , j} 
\ 
\cA_L (l'_2, l'_1, i', \ldots , j') \, \cA_R (-l'_1, -l'_2, (j+1)', \ldots , (i-1)') 
\ , 
\eeq
where the prime means that the momenta have been replaced by their
conformally varied expressions (we also freely replace the loop momenta
and the measure by conformally varied expressions). 
We recall that momenta are written as differences of region momenta as 
\beq
p_i \ := \ x_i - x_{i+1} \ , 
\eeq
and that under an infinitesimal special conformal transformation one has 
\beqa
\label{specialconf}
K^\nu x^\mu &:= & \eta^{\mu \nu} x^2 - 2 \, x^\mu x^\nu
\ . 
\eeqa
In spinor notation, \eqref{specialconf} becomes 
\begin{equation}
  K^{\beta \dot \beta} x_i^{\alpha \dot \alpha} = -2 \, x_i^{\alpha \dot
    \beta} \, x_i^{\beta \dot \alpha}
    \ , 
\end{equation}
and one can easily check that the transformations
\begin{align}
  K^{\beta \dot \beta}\lambda_i^\alpha &= -2\,  (x_i^{\alpha \dot \beta}\lambda_i^\beta  \, -\, 
  \mu_i \lambda_i^\alpha \lambda_i^\beta \tilde \lambda_i^{\dot
    \beta})\ , \\
  K^{\beta \dot \beta}\tilde \lambda_i^{\dot \alpha} &= -2\,   (x_i^{\dot \alpha  \beta} \lt_i^{\dot\beta}\, -\, 
  (1-\mu_i) \tilde \lambda_i^{\dot \alpha} \tilde\lambda_i^{\dot
    \beta} \lambda_i^{ \beta} )
    \ , 
\end{align}
are consistent with this 
(consider $ K^{\beta \dot \beta }
( 
\lambda_i^\alpha \tilde \lambda_i^{\dot \alpha} ) = K^{\beta \dot \beta }
 x_{i \, i+1} ^{\alpha \dot \alpha}$).
 The standard choice of spinor transformation \cite{dhks} is obtained by
simply setting $\mu_i=0$. We will have to consider these
general 
transformations involving the free parameter $\mu_i$ in order to
correctly cope with the transformation of the spinors associated with
the loop momenta. 

We have
\begin{equation}
K^\mu x_{ii+2}^2 \ = \ -2(x_i^\mu  + x_{i+2}^\mu )x_{ii+2}^2 \ , 
\end{equation}
and
\begin{align}
K^\mu \lan i i+1 \ran \ &= \ -2 \Big[x_i^\mu (1- \mu_i) + x_{i+1}^\mu
(\mu_i  - \mu_{i+1}) + x_{i+2}^\mu \mu_{i+1}\Big]  \lan i i+1 \ran\ ,  \\
K^\mu [i i+1 ] \ &= \ -2 \Big[x_i^\mu \mu_i + x_{i+1}^\mu
(-\mu_i  + \mu_{i+1}) + x_{i+2}^\mu (1-\mu_{i+1})\Big] [ i i+1 ]
\ .
\end{align}
After  performing the fermionic  integrations and one of the two momentum
integrations, as well as a shift in the integration variable to $y= l_1 +
x_{i}$, the cut superamplitude becomes
\beq
\int d^D y \, \delta^{(+)}(l_1^2) \, \delta^{(+)} (l_2^2) \ \langle l_1 l_2
\rangle^4 \cA_L \cA_R 
\ , 
\eeq
where $l_1=\lambda_{l_1} \tilde \lambda_{l_1} = y-x_i$ and $l_2=\lambda_{l_2}
\tilde \lambda_{l_2} = x_{j+1}-y$. 

The transformation of the various components of the cut
diagram read as follows, 
\begin{align}\nonumber
&K \cA_L(l_2, l_1, i, \ldots , j) \\\nonumber
=& \ 2 \Big[ \sum_{a=i+1}^j x_a (1- 2\mu_a
+2 \mu_{a-1}) + x_{j+1}(1- 2\mu_{l_2}+2\mu_j)  \\
&\qquad + y (1-2\mu_{l_1}+2
\mu_{l_2}) + x_i(1-2
\mu_i + 2 \mu_{l_1} )  \Big] \cA_L(l_2,
l_1, i, \ldots , j) \ , 
\end{align}
and 
\begin{align}
&K \cA_R(-l_1, -l_2, j+1, \ldots , i-1)\nonumber\\
=&  \ 2 \Big[ \sum_{b=j+2}^{i-1} x_b (1- 2\mu_b
+2 \mu_{b-1}) + x_{i}(1- 2\mu'_{l_1}+2\mu_{i-1})\nonumber  \\
&\qquad + y (1-2\mu'_{l_2}+2
\mu'_{l_1}) + x_{j+1}(1-2
\mu_{j+1} + 2 \mu'_{l_2} )  \Big]
\cA_R(-l_1, -l_2, j+1, \ldots , i-1) 
\ . 
\end{align}
Furthermore, one has 
\beqa
K\,  \delta^{(+)}(l_1^2)&=& 2 (x_i+y) \, \delta^{(+)} (l_1^2)\ , \\
K\,  \delta^{(+)}(l_2^2)&=& 2 (x_{j+1}+y) \, \delta^{(+)} (l_2^2)\ , \\
K \, d^D y& =&-2 D\,  y \, d^D y \ . 
\eeqa
Here the $\mu'$ parameters are defined as $\mu'_{l_1}:= 1-\mu_{l_1}$
and $\mu'_{l_2}:=1-\mu_{l_2}$,  accounting for the fact that,  in  $\cA_R$,
$l_1$ and $l_2$ are in the reverse cyclic ordering to $\cA_L$. The
covariance of all tree-level amplitudes under conformal transformations
was proved in \cite{bhtrec} and indeed  covariance under the more
general transformations defined here was also proven there.

Putting all this together we get that the cut amplitude transforms
with weight
\beq
  2 \sum_{a=1}^n x_a^\mu (1-2\mu_a+2\mu_{a-1})+ 2 (4-D) y^\mu \ .
\eeq
The first term is the expected covariant term, whereas the second  is
the contribution to the one-loop dual conformal anomaly quoted in \eqref{anomalycut}.

\end{document}